\documentclass[12pt,fleqn]{article}
\usepackage{verbatim,amssymb,amsmath,graphics,amsthm}

\newcommand{\nb}{\nonumber}

\newcommand{\xbs}{\boldsymbol}
\newcommand{\p}{\hat{p}}
\newcommand{\<}{\langle}
\renewcommand{\>}{\rangle}
\newcommand{\s}{\sigma}
\newcommand{\sm}{\mathsf{s}}

\renewcommand{\d}{\boldsymbol{\p}_{d}\sigma_{d}[\sm_d j_d]\eta_{ef}n_{ef}}
\renewcommand{\c}[1]{\boldsymbol{\p}^{#1}_{c}\sigma_{c}^{#1}[m_c j_c]}
\newcommand{\2}[2]{\boldsymbol{\p}_{#2}^{#1}\sigma_{#2}^{#1}[\sm_{#2}^{#1}
j_{#2}^{#1}]\eta_{#2}^{#1}n_{#2}}
\newcommand{\cef}[2]{\boldsymbol{\p}^{#1}_{#2}\sigma_{#2}^{#1}[\sm_{#2}^{#1}
j_{#2}^{#1}]
\sm_{ef}^{#1}j_{ef}^{#1}\eta_{#2}^{#1}\eta_{ef}^{#1}n_{#2}}

\begin{document}
\title{Analysis of resonance production using relativistic Gamow vectors}
\author{H. Kaldass\footnote{hani@physics.utexas.edu}\\
The Arab Academy of Science \& Technology \\ Cairo, Egypt}
\date{} \maketitle
\begin{abstract}
The calculation of an amplitude involving resonance production is presented. This calculation 
employs for the resonance state a relativistic Gamow vector. It is used for investigating
the question of compatibility of the relativistic Gamow vectors kinematics, defined by real
$4$-velocities and complex mass, with the stable particle kinematics; or in other words, the
integration of the Gamow vectors with the conventional Dirac bra-ket formalism. The calculation
demonstrates a consistent framework comprising stable and Gamow vectors.
\end{abstract}
\section{Introduction}
The description of unstable particles with state vectors stems from the 
point of view that unstable particles are not less fundamental than the 
stable ones. In fact, unstable particles are listed along the stable ones in 
the Particle Data Table~\cite{particledata} and attributed values for 
the mass, the spin and the width (or lifetime). Hence a zero value for the width (or an infinite lifetime) 
is what distinguishes a stable from an unstable particle. The relativistic Gamow vectors provide state 
vectors for unstable particles through a precise formulation of complex mass representations of the 
Poincar\'e group with self-adjoint generators~\cite{paper2,paper3}. 
These representations are ``minimally complex'', in the 
sense that, while the mass is complex, the $4$-velocity is real. Furthermore, they have the remarkable feature
of being representations for the causal Poincar\'e semigroup into the forward light cone, hence an exponential
decay law defined only in the forward light cone.

The relativistic Gamow vectors have all the properties that are expected for state
vectors describing unstable particles. We explore in what follows the kinematical implications
of the characteristic properties of the relativistic Gamow vectors, namely those of real $4$-velocities
and complex mass. In other words, the question of integrating these properties 
with the standard kinematics of stable particles into a consistent framework is addressed.
The investigation is carried here by analyzing an example of resonance production.

\section{Resonance Production}
We consider the production of an unstable particle $d$ in the
reaction
\begin{equation}
\label{reaction1}
a+b\rightarrow c+d\,.
\end{equation}
We restrict ourselves further to a specific channel
where $d$ decays into two particles $e$ and $f$:
\begin{equation}
\label{reaction2}
d\rightarrow e+f\,.
\end{equation}
In the event that $d$ is a stable particle, the center of mass energy
square of the reaction, $\sm_{ab}$($=(p_a+p_b)^2$), is 
\begin{equation}
\label{sab}
\sm_{ab}=(p_c+p_d)^{2}\,.
\end{equation}
If $d$ is an unstable particle with the attributes of a complex mass
$\sqrt{\sm_d}$,
\begin{equation}
\sqrt{\sm_d} = M_R-i\frac{\Gamma_R}{2}\,,
\end{equation}
 and real $4$-velocity $\p_d$, then the momenta is complex 
$p_d=\sqrt{\sm_d}\p_d$. Therefore,~\eqref{sab} is not satisfied. To find
out the result corresponding to~\eqref{sab} in the complex mass case,
we consider the bra-ket involving directly the variables $p_c$, $p_d$ 
and $\sm_{ab}$:
\begin{equation}
\label{braket}
\<\,[d][c]^{-}\,|\,[ab]^{+}\,\> 
\equiv \langle\d,\c{}^{-}|\2{}{ab}^{+}\rangle\,.
\end{equation}
In \eqref{braket}, 
\begin{itemize}
\item $|\,[ab]^{+}\,\> \equiv |\2{}{ab}^{+}\rangle\,$
is a $4$-velocity basis vector for the in-states of the $a$ and $b$ particles, diagonal in the total mass-square 
$\sm_{ab}$ and angular momentum $j_{ab}$ of the $[ab]$~\footnote{\label{fn}$[ab]$ denotes an irreducible
representation space of the two-particle system, $a$ and $b$. Likewise, $[cef]$ denotes an irreducible
representation space for the three-particle system, $c$, $e$ and $f$.} system~\cite{hani}. 
It results from the reduction of the direct
product space of the $a$ and $b$ particles, $[a]\otimes[b]$, into a direct sum of irreducible representation
spaces~\cite{wightman,joos,macfarlane}. It is labeled by the space component $\xbs{\p}_{ab}$ of the $4$-velocity 
$\p_{ab}=(p_a+p_b)/\sqrt{\sm_{ab}}$, the three-component $\s_{ab}$ of $j_{ab}$, degeneracy quantum numbers $\eta_{ab}$ 
that consist of the orbital angular momentum $l_{ab}$ and the total spin $s_{ab}$ of 
the $[ab]$ system: $\eta_{ab} = \{l_{ab},s_{ab}\}$,
and particle species label $n_{ab}$ designating the masses and spins of $a$ and $b$: $n_{ab} = \{m_a,s_a,m_b,s_b\}$.
\item $|\,[c]^-\,\>\equiv |\c{}^{-}\>$ is a $4$-velocity basis vector for the $c$ particle.
\item $|\,[d]^-\,\>\equiv |\d^-\>$ is a relativistic Gamow vector associated with the $[ef]$ decay channel. It is 
characterized by a complex mass square $\sm_d$, and real $4$-velocity $\p_d$.
\end{itemize}
If $d$ is a stable particle,~\eqref{braket} is an $S$-matrix element for the 
transition $ab\rightarrow cd$ for a certain partial-wave of $ab$.
In this case,~\eqref{braket} is proportional to the delta functions
giving rise to the conservation condition~\eqref{sab}:
\begin{multline}
\label{conservation}
\langle\xbs{\p}_{d}\s_{d}[\sm_d j_{d}],\c{}^{-}|\2{}{ab}^{+}\rangle =\\
\delta(\xbs{p}_{ab}-\xbs{p}_{c}-\xbs{p}_{d})
\delta(\sm_{ab}-(p_d+p_c)^{2})\\
\langle\!\langle
\xbs{\p}_{d}\s_{d}[\sm_d j_{d}],
\c{}|S|\2{}{ab}\rangle\!\rangle\,.
\end{multline}
In~\eqref{conservation}, $\<\!\<\cdots|S|\cdots\>\!\>$ is a reduced
$S$-matrix element resulting from the factoring out of the conservation 
delta-functions.
We calculate how the right side of~\eqref{conservation} gets
modified in the event that $d$ is unstable. For this 
calculation, we use for $d$ a Gamow vector.

A calculation of \eqref{braket} with a Gamow vector is feasible by virtue of the integral representation~\cite{paper2}:
\begin{equation}
\label{gv}
|[d]^-\>\equiv|\d^{\quad-}\rangle=\frac{i}{2\pi}\int_{-\infty_{II}}^{\infty} 
ds{''}{}\frac{|\xbs{\p}_{d}\s_d[\sm''j_d]
\eta_{ef}n_{ef}^{\quad-}\>}{\sm''-\sm_d}\,.
\end{equation}
We should mention that the equality in \eqref{gv} is not postulated, 
but is rigorously derived~\cite{paper2} 
in a relativistic Rigged Hilbert Space~\cite{gelfand} scattering theory~\cite{gv,bg}
from the definition of an unstable particle as being the quantity associated with the complex poles
of the partial $S$-matrix.

\section{Calculation of \eqref{braket} with $d$ unstable}
In order to calculate~\eqref{braket} with the use of the integral
representation~\eqref{gv}, we need to insert in~\eqref{braket} two
complete set of basis vectors. A complete set of basis vectors is a
resolution of the identity with the general form:
\begin{eqnarray}
\label{resolution}
I&=&\sum |\text{one particle basis}\>\<\text{one particle basis}|\nonumber\\
&+&\sum|\text{two particles basis}\>\<\text{two particles basis}|+\cdots\,.
\end{eqnarray}
In~\eqref{resolution}, the summation is over all particle species labels and 
the physical ranges of all quantum numbers, with the understanding that
the summation refers to an integration with a specific measure for the 
continuous quantum numbers. 
The layout for the calculation of \eqref{braket} is as follows :
\begin{eqnarray}
\lefteqn{\langle\,\,[d][c]^-\,\,|\,\,[ab]^+\,\,\rangle}\nb\\
&&=\sum_{[ef][c]}\langle\,\,[d][c]^-\,\,|\,\,[ef][c]^-\,\,\rangle
\langle\,\,[ef][c]^-\,\,|[ab]^+\,\,\rangle\label{first}\\
&&\!\!\!\!\!\!\!\!\!\!\!\!\!\!\!\!=\sum_{[ef][c]}\sum_{[cef]}
\underbrace{\langle\,\,[d][c]^-\,\,|\,\,[ef][c]^-\,\,
\rangle}_{\text{Integral representation}}
\underbrace{\langle\,\,[ef][c]^-\,\,|\,\,[cef]^-\,\,\rangle}_{\text{C-G coefficient}}\underbrace{
\langle\,\,[cef]^-\,\,|\,\,[ab]^+\,\,\rangle}_{\text{$S$-matrix element}}
\,.\label{layout}
\end{eqnarray}
The above equation means that
we first, in \eqref{first}, insert a complete set of out basis vectors.
From~\eqref{resolution}, we retain only the basis vectors with
particle species quantum numbers matching those of $\<\,[d][c]^-|$.
These are the $[ef]\otimes[c]$ basis system which carry the
particle species labels $n_{ef}$ and $n_{c}$,
Then, in \eqref{layout}, we reinsert 
the identity~\eqref{resolution}, and in this case we retain 
the out basis vectors of the $[cef]$ $^\text{\ref{fn}}$ system, $|[cef]^{-}\>$, 
which naturally carries the same particle species labels as in the  
$|[ef][c]^{-}\>$ kets.
In \eqref{layout}, the first term is evaluated
using the integral representation \eqref{gv} of the Gamow vector.
The second term is the Clebsh-Gordan coefficient that arises 
from the change of basis from the direct product basis vectors
$|[ef][c]^-\>$ into the direct sum basis vectors $|[cef]^-\>$~\cite{macfarlane};
and the third term is an $S$-matrix element for the transition
$ab\rightarrow cef$, which is diagonal in the total angular momentum and
mass-square (rest-mass energy square) of the in-state, $\sm_{ab}$.

Upon performing the calculation outlined in \eqref{layout}, the result is:
\begin{multline}
\label{unstablecase}
\langle\d,\c{}^{-}|\2{}{ab}^{+}\rangle\\
\!\!\!\!\!\!=\frac{-i}{2\pi}                      
\frac{1}{\sm'_{ef}-\sm_{d}^{*}}
\frac{1}{\sm_{ab}^{3/2}}\frac{1}{\sqrt{2\sqrt{\sm'_{ef}}}}
N(\sm_{ab},\sm_{ef}',m_c^2)2\p_{ab}^{0}
\delta
\left(\xbs{\p}_{ab}-\frac{\sqrt{\sm_{ef}'}\xbs{\p}_{d}+\xbs{p}_{c}}
{\sqrt{\sm_{ab}}}\right)\frac{x}{x+\p_{d}.p_{c}}\\
\times
\sum_{\eta_{o}'}
P(p_{ef}'p_{c},\s_{d}\s_{c},j_{ab}\s_{ab}\eta_{o}')
\<\sm_{ef}'j_d\eta_{o}'\eta_{ef}n_{o}|S^{j_{ab}}(\sm_{ab})|
\eta_{ab}n_{ab}\>\,.
\end{multline}
where
\begin{itemize}
\item $\displaystyle\sm_{ef}'=x^{2}=(p_{ab}-p_{c})^{2}$,
\item the subscript $o$ in $\eta_o'$ and $n_o$ refers to the label: $cef$,
\item $\<\sm_{ef}'j_d\eta_{o}'\eta_{ef}n_{o}|S^{j_{ab}}(\sm_{ab})|\eta_{ab}n_{ab}\>$ is a reduced $S$-matrix element:
\begin{multline}
\label{2}
\langle\cef{'}{o}^{-}|\2{}{ab}^{+}\rangle=\\
\!\!\! 2\p_{ab}^{0}\delta(\boldsymbol{\p}_{ab}-\boldsymbol{\p}^{'}_{o})
\delta(\sm_{ab}-\sm_{o}')
\delta_{j_{o}'j_{ab}}\delta_{\sigma_{o}'\sigma_{ab}}
\langle\sm_{ef}'j_{ef}'\eta_{o}'\eta_{ef}'n_o
|S^{j_{ab}}(\sm_{ab})|\eta_{ab}n_{ab}\rangle\,,
\end{multline}
\item $N$ is a normalization factor, and $P$ is a factor involving rotation matrices and Clebsh-Gordan
coefficients~\cite{wightman,macfarlane,hani}.
\end{itemize}
The limit of \eqref{unstablecase} as $\sm_d$ becomes real is found to be: 
\begin{multline}
\label{stablecase2}
\<\d,\c{}^{-}|\2{}{ab}^{+}\>\\
\stackrel{\Gamma_d\rightarrow 0}{=}
\frac{1}{\sm_{ab}^{3/2}}\frac{1}{\sqrt{2\sqrt{\sm_{d}}}}N(\sm_{ab},\sm_{d},m_{c}^{2})
2\p^{0}_{ab}\delta(\xbs{\p}_{ab}-\xbs{\p}_{o}^{'})
\frac{x}{x+\p_d.p_c}\delta(\sm_d-(p_{ab}-p_{c})^{2})\\
\times\sum_{\eta_{o}'}P(p_{d}p_{c},\s_{d}\s_{c},j_{ab}\s_{ab}\eta_{o}')
\<\sm_{d}j_d\eta_{o}'\eta_{ef}n_{o}|S^{j_{ab}}(\sm_{ab})|
\eta_{ab}n_{ab}\>\,.
\end{multline}
where 
\begin{equation}
\nb
\xbs{\p}_{o}^{'}=\frac{\sqrt{\sm_{d}}\xbs{\p}_{d}+\xbs{p}_{c}}{\sqrt{\sm_{ab}}}
\end{equation}
\begin{equation}
\label{sod}
\sm_{o}'=(p_d+p_c)^{2}\,.
\end{equation}
\section{Summary and Discussion}
We observe that the passage from real $\sm_d$ \eqref{stablecase2}
to a complex $\sm_d$ \eqref{unstablecase} occurs with the following changes:
\begin{itemize}
\item The mass-square condition $\sm_d = (p_{ab}-p_{c})^{2}$ \eqref{sab} expressed in \eqref{stablecase2}
by the delta function: 
\begin{equation}
\delta(\sm_d-(p_{ab}-p_{c}))^2
\end{equation}
gets replaced in the unstable case \eqref{unstablecase} by a Breit-Wigner distribution in $(p_{ab}-p_c)^2$:
\begin{equation}
\frac{-i}{2\pi}\frac{1}{\sm'_{ef}-\sm_d^*} = \frac{-i}{2\pi}\frac{1}{(p_{ab}-p_c)^2-(M_R+i\Gamma_R)^2}\,.
\end{equation}
This calculational result can be traced to the fact that an unstable particle, unlike
a stable one, does not have a definite (real) mass, rather is the superposition of the {\it whole} spectrum of masses 
with Breit-Wigner weights, as expressed in \eqref{gv}. 
\item The condition on the space-components of the $4$-velocities in the stable case \eqref{stablecase2}:
\begin{equation}
\xbs{\p}_{ab} = \frac{\sqrt{\sm_d}\xbs{\p}_d+\xbs{p}_c}{\sqrt{\sm_{ab}}}
\end{equation}
expressed by the delta function:
\begin{equation}
\delta(\xbs{\p}_{ab}-\xbs{\p}_o')
\end{equation}
gets replaced by:
\begin{equation}
\xbs{\p}_{ab} =\frac{\sqrt{(p_{ab}-p_c)^2}\xbs{\p}_d+\xbs{p}_c}{\sqrt{\sm_{ab}}}\,.
\end{equation}
Hence, even though $\sm_d$ is complex, $\p_{ab}$ remains real as the quantity 
$\sqrt{\sm_{ef}'}=\sqrt{(p_{ab}-p_{c})^{2}}$ replaces $\sqrt{\sm_{d}}$ in \eqref{stablecase2}, and
$\p_d$ is real according to a defining postulate of the relativistic Gamow vectors.
\end{itemize}
\section*{Acknowledgment}
This work is part of a collaboration with the University of Texas at Austin sponsored
by the US National Science Foundation Award No. OISE-0421936.

\end{document}